\documentclass[11pt]{article}
\usepackage[utf8]{inputenc}
\usepackage[tmargin=1in,bmargin=1in,lmargin=1in,rmargin=1in]{geometry}
\usepackage{graphicx}
\usepackage{float}
\usepackage{amssymb}
\usepackage{adjustbox}
\usepackage{epstopdf}
\usepackage{enumitem}
\usepackage{amsmath}
\usepackage{biblatex}
\usepackage{textcomp}
\usepackage{chngcntr}
\usepackage{comment}
\usepackage{caption}
\usepackage{subcaption}
\usepackage[T1]{fontenc}
\usepackage{bm}
\usepackage{float}

\usepackage{authblk}
\usepackage{multirow}
\usepackage{multicol}
\usepackage{comment}
\usepackage{placeins}
\usepackage{ragged2e}
\allowdisplaybreaks
\usepackage{hyperref}
\hypersetup{
    colorlinks=true,
    linkcolor=blue,
    filecolor=magenta,      
    urlcolor=cyan,
    citecolor=blue,
}
\title{Impacts of social and impact heterogeneity on social-climate outcomes}
 \author[1,2,*]{Amrita Punnavajhala}
 \author[3,4,5,6]{Brian Beckage}
 \author[2]{Madhur Anand}
 \author[7]{Daniele Visioni}
 \author[8]{Jared Farley}
 \author[1]{Chris T. Bauch}

\affil[1]{Department of Applied Mathematics, University of Waterloo, Waterloo, ON N2L 3G1, Canada}
\affil[2]{School of Environmental Sciences, University of Guelph, Guelph, ON N1G 2W1 , Canada}
\affil[3]{Department of Plant Biology, University of Vermont, Burlington, VT 05405, USA}
\affil[4]{Department of Computer Science, University of Vermont, Burlington, VT 05405, USA}
\affil[5]{Gund Institute for Environment, University of Vermont, Burlington, VT 05405, USA}
\affil[6]{Vermont Complex Systems Institute, University of Vermont, Burlington, VT 05405, USA}
\affil[7]{Department of Earth and Atmospheric Sciences, Cornell University, Ithaca, NY 14850, USA}
\affil[8]{Sibley School of Mechanical and Aerospace Engineering, Cornell University, Ithaca, NY 14850, USA}
\affil[*]{Corresponding author: apunnavajhala@uwaterloo.ca}
\begin{document}

\maketitle
\begin{justify}

\section*{Abstract}
Regional heterogeneity in social characteristics, temperature change, and vulnerability to climate impacts is likely to influence the magnitude of anthropogenic climate change, but has not been considered in coupled social-climate models, which seek to represent interactions between social and climate dynamics.  Here, we examine how the projected mean global temperature anomaly and population support for mitigation respond to heterogeneity in these factors across five regions of the world, using a coupled social-climate model. We find that heterogeneity in climate impacts increases the temperature anomaly by 0.2$^\circ$C, while social heterogeneity increases it by an additional 0.1$^\circ$C.  The projected temperature anomaly also increases with higher variability in climate impacts across regions, even for the same average global climate impact. Finally, we identify a social-climate tipping point, where low vulnerability to impacts under existing social conditions in one region can tip the system into an alternative slow-mitigation, high-temperature state. Our results show that heterogeneity in climate impacts leads to higher global mean temperatures  and efforts to reduce global disparities could improve both social and climate outcomes.

\end{justify}

\section*{Introduction}

The Earth is  $\sim$1.3$^\circ$C hotter than it was before the Industrial Revolution, and the impacts of such rapid warming are increasingly apparent \cite{mora2018broad,hughes2018spatial,burrows2024systematic,lehmann2020complex}. The negative effects of continued anthropogenic climate change will affect every aspect of our lives and livelihoods \cite{IPCC2023SPM}. The mitigation of climate change  is seemingly straightforward: reduce, and, ideally, eliminate greenhouse gas emissions \cite{Dessler2021Climate}. However, despite the threat of future climate change, a collective global effort to mitigate remains elusive \cite{LancetCountdown2025}. There are many reasons for this lack of progress on addressing anthropogenic climate change, varying across scales and regions, from government level concerns about equity \cite{holz2023tempering} and mitigation costs \cite{koberle2021cost}, to individual psychology \cite{mccright2016ideology} and climate skepticism \cite{biddlestone2022climate}, among others. For mitigation to succeed, it is important, though challenging, to understand these underlying processes that inhibit a coherent human response to a potentially existential threat to our society.

Mathematical models can help provide insight into dynamics of complex social-ecological systems, for example, how raising awareness can help coral reef conservation \cite{thampi2018socio} or how homophily can slow down vaccine uptake \cite{he2024effect}. Much climate change research studies nonlinear feedback effects between geophysical processes \cite{armstrong2022exceeding}, or feedbacks governing individual mitigation behaviour \cite{hampton2023choices, bretter2025public}. But there is little work studying feedback between behaviour and geophysical processes, despite evidence showing the influence of climate change impacts on mitigation support \cite{andre2024globally, cologna2025extreme}, and this, in turn, influences the state of climate change \cite{schaffer2022policymakers, gazmararian2025public}. Integrated assessment models (IAMs), which are widely used to inform policy, combine socioeconomics with climate models \cite{riahi2017shared}, but most of them lack a dynamic behavioural or social component (exceptions to this are \cite{lamperti2018faraway} and \cite{lackner2025opinion}, for example). Coupled social-climate models \cite{beckage2018linking, bury2019charting, moore2022determinants, donges2020earth}, on the other hand, represent social dynamics endogenously.  However, they are at early stages of development and make many simplifying but unrealistic assumptions. One such assumption is that global populations are homogeneous, whereas the representation of socio-cultural heterogeneity as well as heterogeneity in temperature changes across regions could lead to more complex and realistic projections and understanding of anthropogenic climate change \cite{menard2021conflicts, punnavajhala2025implications}. However, it is not clear how modelling heterogeneous populations alters results relative to a homogeneous population assumption, since there is no comparison of these social-climate outcomes within a single modelling framework.

In a laboratory setting, a variable of interest can be turned `on' or `off' to study its effects on the system.  But this is not possible with real-world social-cultural heterogeneity, leaving only the option of studying the effects of adding or removing such heterogeneity through model simulations.  Our goal in this work is to use social-climate modelling to understand how social and climate heterogeneity influence the trajectory and dynamics of the coupled social-climate system.  Specifically, we are interested in identifying outcomes  that might not emerge under the assumption of socio-cultural homogeneity, and in understanding the mechanisms that drive system behavior. To this end, we extend an existing social-climate model with empirically backed social dynamics across five regions \cite{punnavajhala2025implications} to include region-specific warming and vulnerability to climate change impacts. Our model is not intended to be predictive but rather to lead to a better understanding of what outcomes could emerge from empirically plausible scenarios that include socio-cultural and climate impact heterogeneity.

\begin{figure}%[tbhp]
\centering
\includegraphics[width = 0.95\linewidth]{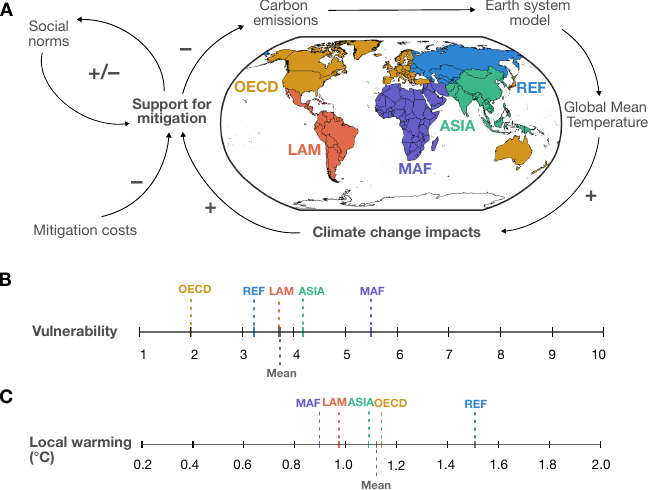}
\caption{Model overview. ($A$) overview of the coupled social-climate model with feedback between different model components. Social dynamics are local to each of the five regions shown in the map. ($B$) Vulnerability to climate change impacts in each region measured as a number between 0 and 10 with a higher number representing greater vulnerability. ($C$) Warming (in $^\circ$C) experienced by each region for 1$^\circ$C global warming (`local warming'), relative to pre-industrial. (See Methods).}
\label{fig:model_diagram}
\end{figure}

We model social dynamics in five regions of the world: Asia (ASIA), Latin America (LAM), the Middle-east and Africa (MAF), Organization for Economic Co-operation and Development (OECD) countries and the Reforming Economies of Eastern Europe and the former Soviet Union (REF) (Fig.~\ref{fig:model_diagram}$A$), as defined in the Shared Socioeconomic Pathways (SSP) five region aggregation \cite{bauer2017shared}. This representation captures some degree of real world heterogeneity but is simple enough for detailed analysis of social-climate dynamics in each region. 

The greenhouse gas emissions by each region are assumed to depend inversely on support for climate change mitigation in the population, which in turn is influenced by social conditions and global warming effects felt locally. Each region starts out with a fixed population fraction of mitigation supporters (initial fraction of `mitigators' in our model equations). These individuals, through interactions with one another (`social learning'), either convince others to support mitigation, or are convinced to stand against mitigation--thus becoming `non-mitigators'--depending on which stance appears to provide a greater benefit (`utility'). The utility of mitigation depends on the economic cost of mitigation, the cost of climate change impacts, and the strength of social norms that reinforce prevailing majority opinion, whether that be mitigation or non-mitigation. Mitigation across all regions influences the quantity of anthropogenic carbon emissions entering the atmosphere. Atmospheric carbon is partially absorbed into land and ocean sinks. The remainder determines net solar radiation at the earth's surface, and thus, the global mean surface temperature. The mean temperature anomaly (relative to pre-industrial) is then mapped to the five regions, so they experience unequal warming as well as unequal impacts, the latter driven by differences in socioeconomic vulnerability. These, combined with region-specific social and economic processes, affect mitigation support, forming a feedback loop between the social and climate sub-systems.  Full details of the model appear in the {Methods}.

Since mitigators are initially in a minority in all regions, social norms discourage people from supporting mitigation. However, the suppression of mitigation drives up global warming and its impacts which, together with projected economic benefits of clean energy, can outweigh norms to flip the population over to supporting mitigation instead. Mitigation within a region is also influenced by other regions through feedback from shared global warming impacts. If one region cuts carbon emissions, this lessens global warming impacts for all regions, thus reducing the incentive for other regions to mitigate. Conversely, if a region is slow to mitigate, this could worsen impacts on other regions thus driving them to mitigate sooner.

 We study the influence of socio-cultural and climate change impact heterogeneity on mitigation support and the mean global surface temperature anomaly.  Our framework for analysis is to use the model as a virtual laboratory: by adding/removing heterogeneity while keeping all other model conditions unchanged, we can conduct a virtual experiment using counterfactual worlds that do not have heterogeneity, to understand how heterogeneity influences temperature projections and population support for climate change at the regional level. 

We consider heterogeneity in `social' factors, that is, the net cost of mitigation $\beta(t)$, strength of social norms ($\delta$) and initial mitigation support in each region ($x_0$), and `impact' parameters, vulnerability to impacts ($v$) and location specific warming experienced for $1^\circ$C global warming (`local warming', $m$). The vulnerability to impacts, $v$, is a number between 0 and 10, with a higher values corresponding to greater vulnerability to climate change impacts. We simulate four scenarios that outline the spectrum of possibilities (Fig.~\ref{fig:time_series}):

\begin{itemize}
    \item case i: all regions kept at the global average social and impact conditions 
    \item case ii: all regions kept at global average social conditions but retain region-specific impact conditions
    \item case iii: all regions kept at global average impact conditions but retain region-specific social conditions 
    \item case iv: each region is at their region-specific social and impact conditions
\end{itemize}

Next, we examine the effect of introducing variation into a homogeneous world by measuring deviations in the peak temperature compared to the scenario with mean social and impact parameters (case i) by sampling parameters for each region from triangular distributions preserving the mean, but with varying variance (distributions in Fig.~\ref{fig:variance} $A$--$C$, $G$--$I$, $M$--$O$, $S$--$U$). Finally, we examine the role of vulnerability in determining social-climate outcomes in detail. We plot the global peak temperature anomaly as a function of vulnerability ($v$) for each region, varied one at a time (Fig.~\ref{fig:tipping} $A$--$C$), record peak temperature outcomes under 3125 possible integer permutations of vulnerability between 1 and 5 for each region (SI Appendix Fig.~S1 $A$), and study the relationship between vulnerability and peak temperature when sampling vulnerability for each region from a uniform distribution between 0.5 and 5 (SI Appendix Fig.~S2 $A$--$F$).

\section*{Results}

\begin{figure*}[t!]
\centering
\includegraphics[scale=0.9]{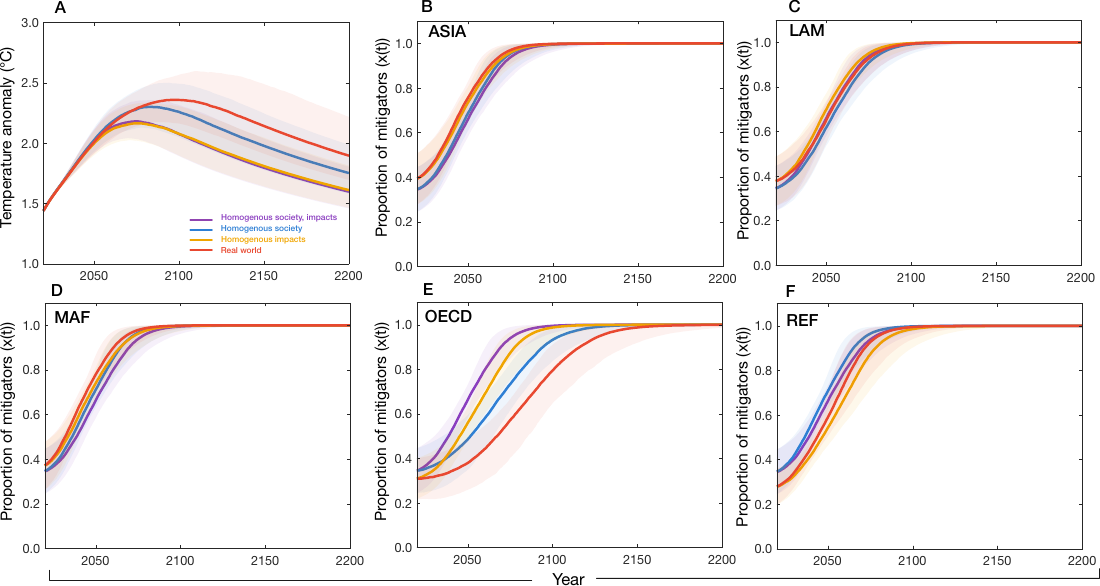}
\caption{Social and environmental factors have differing impacts across regions. The time evolution of the mean global temperature anomaly ($A$) driven by mitigation support under four scenarios in ($B$) Asia (ASIA), ($C$) Latin America (LAM), ($D$) the Middle-east and Africa (MAF), ($E$) Organization for Economic Co-operation and Development (OECD) countries, and ($F$) the Reforming Economies of Eastern Europe and the former Soviet Union (REF). Solid lines represent median trajectories with shading outlining the 0.25 and 0.75 quantiles.}
\label{fig:time_series}
\end{figure*}

Global heterogeneity resulted in the greatest increase in global mean temperature across the four scenarios, with a peak temperature of about 2.4$^\circ$C (case iv).  A mean field model with homogeneous climate impacts (case ii) reduced the peak temperature by approximately 0.2$^\circ$C,  homogeneous socio-cultural conditions reduced temperature by approximately 0.1$^\circ$C (case iii), and the inclusion of homogeneity in both climate impacts and socio-cultural conditions reduced peak temperature by about $0.2^\circ$C  (case i, similar to the case ii) (Fig.~\ref{fig:time_series}$A$). These differences in temperatures across scenarios appear to be driven primarily by two of our regions, i.e., OECD and REF (Fig.~\ref{fig:time_series}$E$,$F$) as support does not vary much across the four scenarios in ASIA, LAM and MAF (Fig.~\ref{fig:time_series}$B$--$D$). In these three regions, there is little  potential for improved mitigation in contrast to the OECD and REF regions. For example, mitigation support in 2050 under homogeneous society and climate impacts (baseline mode, case i), would be more than double compared to case iv for the OECD region (Fig.~ \ref{fig:time_series}$E$). 

\begin{figure*}[t!]
\centering
\includegraphics[scale = 0.9]{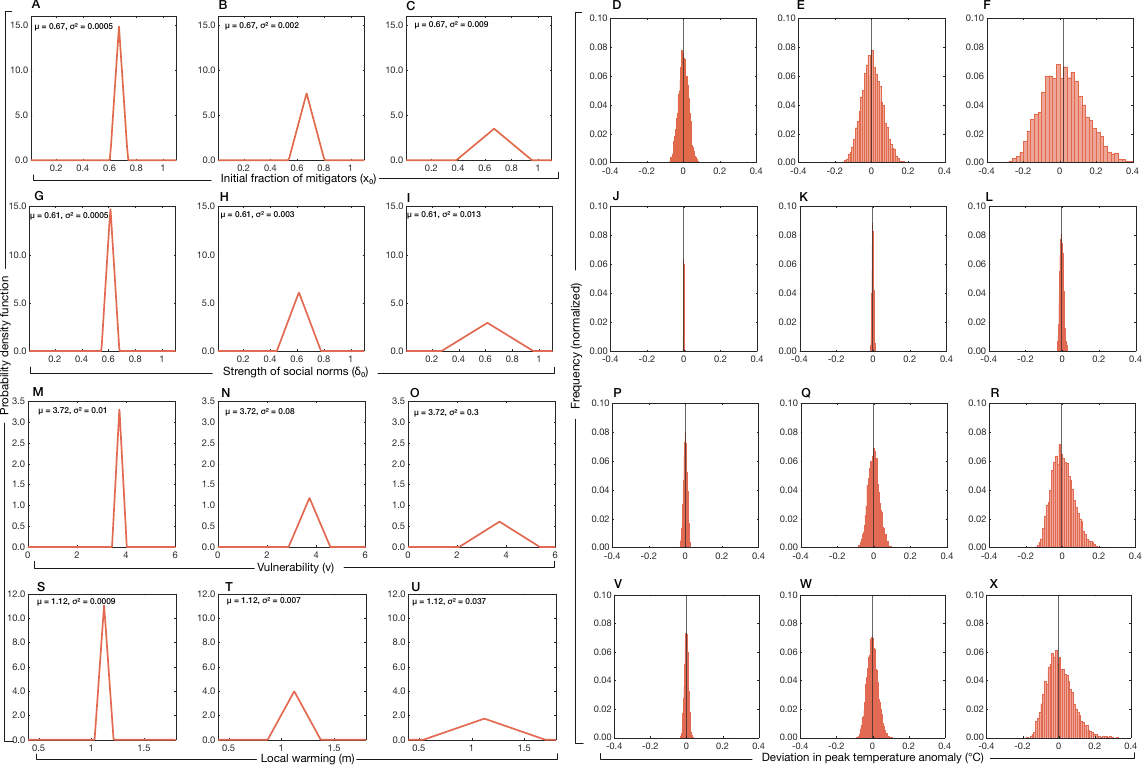}
\caption{Greater variation in parameters causes greater chance of a higher anomaly. Parameter distributions with mean set to global mean, with increasing variance (($A$)--($C$), ($G$)--($I$), ($M$)--($O$), ($S$)--($U$)). Each row corresponds to a different parameter, and variance is increased from left to right. Resulting temperature deviations from the zero variance case are in ($D$)--($F$), ($J$)--($L$), ($P$)--($R$), ($U$)--($X$), with median deviation shown by the vertical line ($N = 5000$).}
\label{fig:variance}
\end{figure*}

Increasing the variance of parameter distributions results in a wider distribution of temperature deviations for all parameters. However all temperatures distributions are positively skewed, which means there is a small probability of a significantly larger temperature anomaly, relative to the non-skewed situation (Fig.~\ref{fig:variance}$D$--$F$, $J$--$L$, $P$--$R$, $U$--$X$). The greatest range of impacts is seen for the initial fraction of mitigators ($x_0$) followed by local warming ($m$) and vulnerability ($v$). Increasing variance within empirically plausible ranges of impact parameters can cause temperature increases in excess of $0.2^\circ$C and decreases of  more than $0.1^\circ$C. Variations in social norms have comparatively smaller effects, impacting the peak temperature by less than $0.1^\circ$C.

\begin{figure*}[t!]
\centering
\includegraphics[width = 0.9\textwidth]{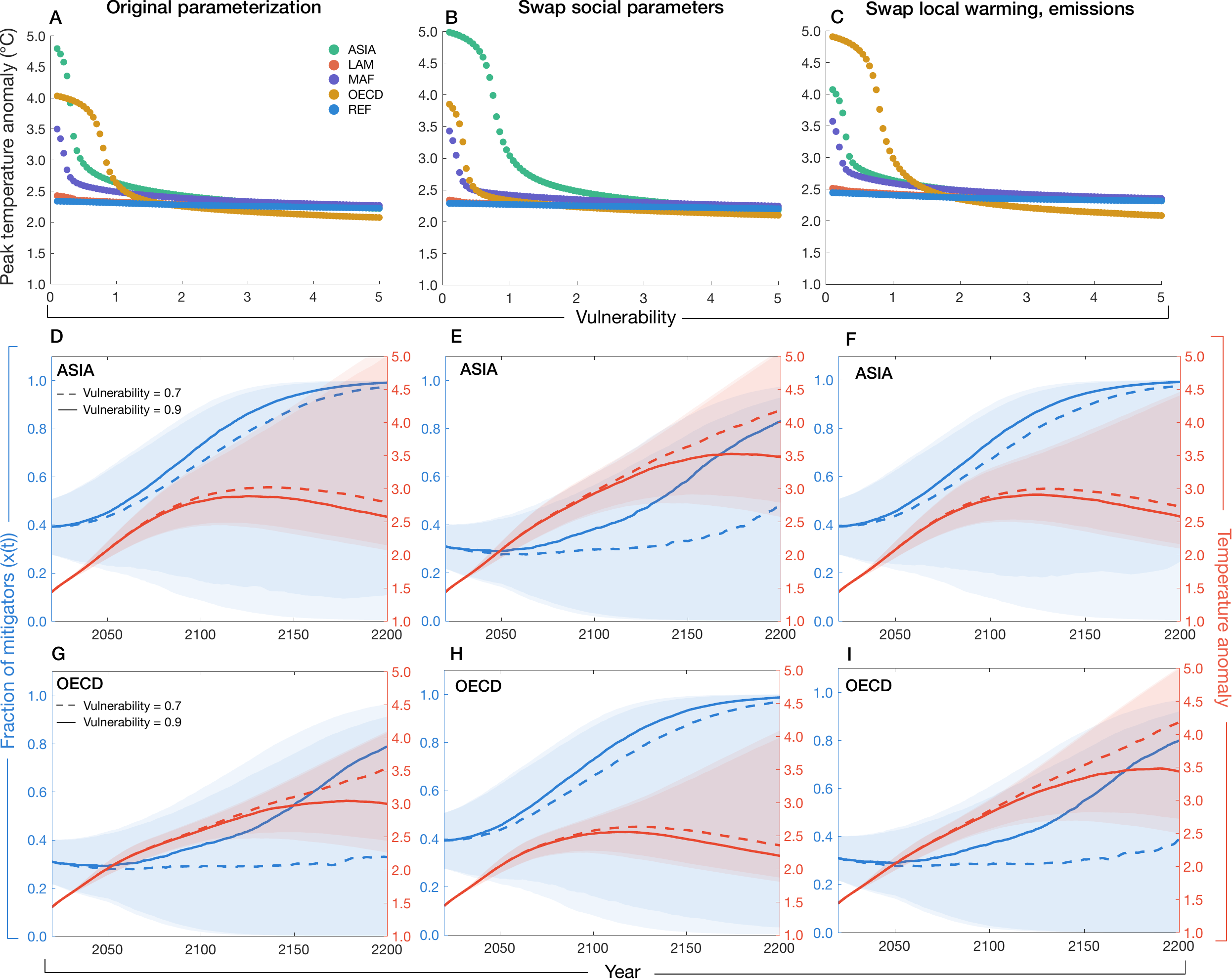}
\caption{Social-climate tipping points are induced at lower vulnerability. ($A$)--($C$) peak temperature anomaly as a function of vulnerability for each region, varied one at a time. Mitigation support (blue, left $y$-axis) and temperature (red, right $y$-axis) time series for different vulnerability levels (dashed for $v = 0.7$, solid for $v = 0.9$) in Asia (ASIA) (($D$)--($F$)) and the Organization for Economic Co-operation and Development (OECD) countries (($G$)--($I$)). (Real world vulnerabilities for ASIA and the OECD are 4.17 and 2.00, respectively). All figures in the first column are using empirical parameter values, the second column is when social parameters are swapped and the third column is for the swapping of emissions and local warming for ASIA and the OECD.}
\label{fig:tipping}
\end{figure*}

 Peak temperature values decrease with increased vulnerability, with ASIA and the OECD causing the largest temperature reductions (Fig.~\ref{fig:tipping}a). There appear to be tipping points, around $v = 0.8$ for the OECD, and near $v=0.2$ for ASIA and $v=0.1$ for MAF. Comparing the time evolution of mitigation support and temperature for $v = 0.7$ and $v = 0.9$ in ASIA and in the OECD, we see a slight difference in the dynamics for ASIA (Fig.~\ref{fig:tipping}$D$), but a more noticeable change in the OECD (Fig.~\ref{fig:tipping}$G$). When $v =0.7$, mitigation support remains more or less constant at around 30\%, but when $v = 0.9$, support ramps up sharply after 2100 because of the increasing temperature (shooting up to about 80\% at the end of the simulation period), and this forces the temperature trajectory down.

To understand what is driving this emergence of this tipping point, we experimented with swapping social parameters for ASIA and the OECD, keeping local warming and emissions intact. This resulted in the two regions swapping tipping points, with ASIA now tipping around $v=0.9$ and the OECD close to  $v = 0.2$ (Fig.~\ref{fig:tipping}$B$). The swapping of social parameters causes mitigation support in ASIA (Fig.~\ref{fig:tipping}$E$) to resemble that originally observed in the OECD (Fig.~\ref{fig:tipping}$G$), and vice-versa (compare Figs.~\ref{fig:tipping}$H$ and \ref{fig:tipping}$D$), suggesting that it is the social parameters for the OECD that are responsible for the tipping point. We further let each region retain their social parameters swapping emissions and local warming instead and observed that tipping points (Fig.~\ref{fig:tipping}$C$) and social dynamics for both regions (Figs.~\ref{fig:tipping}$F$, $I$) resemble the original unswapped case. However, temperature outcomes differ because of the exchange of emissions.

The OECD is the only region whose vulnerability has a clear effect on the peak temperature when varying vulnerability for multiple regions simultaneously. As the OECD gets less vulnerable to climate impacts, the peak temperature rises. Specifically, for all permutations where the OECD has a vulnerability of 1 (lower vulnerability than the real world value of 2), the peak temperature lies above the 7th quantile of all outcomes. There is no discernible pattern between vulnerability in other regions and the peak temperature (SI Appendix Fig.~S1$A$).

All temperatures in the highest quantiles of temperature correspond to slow mitigation in the OECD (SI Appendix Fig.~S1$F$). At low vulnerability, the OECD is less impacted by climate change and because initial mitigation support is low, social norms keep the $x$ trajectory close to the initial condition. Because the OECD is responsible for a substantial proportion of global emissions (SI Appendix, Fig.~S3), this slow mitigation drives temperatures up. ASIA, too, is projected to have a sizeable impact on the temperature through its large emissions, however even with low vulnerability, mitigation in ASIA grows consistently over time. This is because social factors here favour mitigation strongly enough to compensate for any reduction in impacts (SI Appendix Fig.~S1$C$), and similarly for LAM and MAF (SI Appendix Figs.~S1$D$,$E$).

Increased average vulnerability across the world results in lower temperatures.  However there is substantial variation in peak temperature outcomes when mean vulnerability lies between 1 and 4. In fact there appears to be a bifurcation in peak temperatures (SI Appendix Fig.~S2a) with a dense cluster of outcome temperatures between about 2 and 3.4 degrees, and then again between 3.8 and 4.2 degrees. This pattern is seemingly driven by the OECD, where the vulnerability falling below 1 causes a steep increase in peak temperature (SI Appendix Fig.~S2$E$). All regions except the OECD allow either of the two temperature outcomes across all levels of vulnerability (SI Appendix Fig.~S2$B$--$D$,$F$).

ASIA, MAF and the OECD all show the potential to cause tipping points in social-climate dynamics when facing reduced climate change impacts. However the OECD is more likely to cross its tipping point since social parameters do not bolster mitigation support against reductions in impacts, unlike in the other two regions. 

\section*{Discussion}

In our model simulations, heterogeneity in climate impacts and socio-cultural conditions increases the magnitude of global warming by the end of the century, and can induce a bifurcation in social-climate states, driven by differential impacts in the OECD. Heterogeneity has been studied in the context of greenhouse gas emissions \cite{matthews2014national,chancel2022global, hickel2020quantifying}, impacts \cite{waidelich2024climate, kahn2021long}, mitigation support \cite{andre2024globally}, and drivers of mitigation support \cite{andre2024globally, bretter2025public, cologna2025extreme, dablander2025climate}. However, there is little work linking these within a coupled social-climate model.

Bifurcated social-climate outcomes emerge because of the discordance between emissions and impacts in the OECD. When the OECD's vulnerability drops below a threshold, mitigation progress slows down and emissions remain unabated. Because OECD emissions play a significant role in the determination of the global temperature, their inaction causes the temperature to continue rising irrespective of mitigation by other regions. However the scenario is not entirely bleak. We find that social processes could work to reduce temperature change. ASIA, too, has a substantial influence on the global temperature but it is less likely to tip into a slow-mitigation, high-temperature outcome than the OECD because social conditions counteract other factors working against mitigation. It is possible that mitigation starts to gain social acceptance across the OECD given renewable energy already costs less than fossil fuel derived in many places \cite{lorenczik2020projected, nijsse2023momentum} and the behavioural science interventions to promote mitigation are gaining momentum \cite{nielsen2024realizing}.

Our results also reiterate broader concerns about climate change and equity. The OECD countries face reduced climate change impacts, in part, due to their advanced socioeconomic development. However, this has been powered by fossil fuel burning that has led to present-day global warming. On the other hand regions in the Global South (ASIA, LAM and MAF in our model) have contributed little to past fossil fuel burning and are more vulnerable to climate change impacts \cite{fanning2023compensation}. Moreover, countries in the Global South are warmer on average so a degree increase in temperature will cause more economic damage there than it would in colder countries \cite{burke2015global, callahan2022globally}, something we do not explicitly account for in our model. This could further widen the gap between the Global North and South. Another possibility is that climate change adaptation further reduces the OECD's vulnerability \cite{berrang2021systematic}. In the absence of sufficient socioeconomic change, this could not only catapult us into a high temperature outcome but also create a dangerous reinforcing feedback loop: adaptation offers protection from climate change impacts, reducing the incentive to mitigate and developing a `sociocultural niche' \cite{ellis2024anthropocene} that disregards climate change mitigation. While adaptation is necessary and useful, unequal adaptation could worsen impacts in already-vulnerable parts of the world.

This study makes many simplifying assumptions to facilitate the exploration of model dynamics. These include the five-region aggregation which obscures within-region heterogeneity, for example differences between opinions in the Nordic countries and the US. We also restrict social dynamics to each region, although cross-region learning is more realistic. Finally, the replicator dynamics framework represents social processes in a general way and does not allow distinguishing processes at a finer scale, such as mass media effects versus social media effects.

Our investigation into the roles of social conditions and impacts on coupled social-climate outcomes finds that heterogeneity on the whole leads to worse outcomes. However, heterogeneity is an inescapable reality of our world. Recent climate negotiations are working towards reducing inequalities \cite{COP29Fund2023} and this could lead to less severe warming. An important takeaway is that there is a need to represent real world complexity and diversity within social-climate models, because this can throw light on unexpected possibilities, and that pursuing goals like reducing inequalities could benefit both social and climate systems.

\section*{Methods}

 The coupled social-climate model consists of the following differential equations: 

\begin{align}
      \frac{dx_i}{dt} &= \kappa_{0_i} x_i (1-x_i) [- \beta_i(t)   + c_{0_i}  f_i(T_i) + \delta_{0_i} \delta_i (2x_i-1)] \label{x_eqn}\\
       \frac{dC_{at}}{dt} &=  \sum_i (1-x_i)\epsilon_i(t) - P(C_{at},T) + R_{veg} + R_{so} -F_{oc}\label{C_at}\\
    \frac{dC_{oc}}{dt} &= F_{oc}(T,C_{at},C_{oc})\label{C_oc}\\
    \frac{dC_{veg}}{dt} &= P(C_{at},T) - R_{veg}(T,C_{veg}) - L(C_{veg})\label{C_veg}\\
    \frac{dC_{so}}{dt} &= L(C_{veg}) - R_{so}(T,C_{so})\label{C_so}\\
    \frac{dT}{dt} &= \frac{a_E[(F_d(C_{at},T) -\sigma(T+T_0)^4]}{C}\label{T}
\end{align}
This system of equations was adapted from \cite{punnavajhala2025implications} to include location-specific warming and more extensive analysis of heterogeneity impacts. Equation (\ref{x_eqn}) uses replicator dynamics \cite{hofbauer1998evolutionary} to model the evolution of mitigation support over time through social learning. The variable $x_i(t)$ denotes the fraction of the population that supports mitigation in region $i$ at time $t$ and is driven by the social learning rate, $\kappa_{0_i}$, net cost to mitigate, $\beta_i(t)$, cost of climate change impacts, $f_i(T_i)$, and strength of social norms, $\delta_i$, as well as the existing level of mitigation support in the population. The terms $c_{0_i}$ and $\delta_{0_i}$ are scalings used to make model variables dimensionally consistent (so that all terms of the utility function are expressed in terms of the same `currency'). The social learning rate $\kappa_{0_i}$ represents the frequency of interactions between individuals, the net cost to mitigate $(\beta_i - k)$ is based on the costs of renewable energy scaled by energy demand, relative to fossil fuel costs. The cost of climate change impacts $f_i(T_i)$ is a function of warming experienced in each region (`local warming', $m_i$) and vulnerability to impacts $v_i$, and the strength of social norms $\delta_i$ is based on survey responses on the perceptions of mitigation support in each region. Empirical estimates for the initial fraction of supporters in each region ($x_{0_i}$), strength of social norms ($\delta$), local warming ($m$) and vulnerability ($v$) for each region are in SI Appendix, Table S1. 

Equations (\ref{C_at}) - (\ref{T}) represent the climate model with ordinary differential equations for carbon in the atmosphere, $C_{at}$, ocean, $C_{oc}$, vegetation, $C_{veg}$, and soil, $C_{so}$, and the global mean surface temperature anomaly, $T$. These equations include various sources and sinks of carbon: anthropogenic emissions attributed to non-mitigators, $(1-x_i)\epsilon_i(t)$, carbon used for photosynthesis, $P(C_{at},T)$, carbon released through respiration from vegetation and soil, $R_{veg}$ and $R_{so}$, respectively, the flux of carbon from the ocean, $F_{oc}$, and carbon released through plant decay, $L(C_{veg})$. In Equation (\ref{T}), $F_d$ denotes the net downward flux of radiation as influenced by the quantity of carbon dioxide in the atmosphere. Descriptions of functional forms and parameters used for these processes are in SI Appendix, Tables S2 and S3, respectively.

\subsection*{Parameterization}

% \paragraph{Net cost to mitigate $\beta(t)$}
Renewable energy cost forecasts are from the `fast transition' scenario of \cite{way2022empirically}, which assumes the deployment of renewable energy continues at historical rates.  These are multiplied by energy demand forecasts for each region under SSP 2 from \cite{bauer2017shared}, giving us $\beta_i(t)$. From this we subtracted fossil fuel costs based on estimates for coal-powered LCOEs from \cite{way2022empirically}, assumed to be constant over time, and multiplied by the global average energy demand under SSP2. Projected renewable and fossil fuel costs are in Appendix SI Fig.~S4.

We used scalings from \cite{farley2026climate} to calculate the local warming coefficient $m_i$ by mapping the global temperature anomaly to temperature anomalies at specific locations. We multiplied the global temperature anomaly by the scaling for each latitude-longitude coordinate, to get the corresponding `local' temperature anomaly at that coordinate. We then calculated area-weighted averages of these scalings across grid points within each country, and used country scalings to calculate an area-weighted scaling for each region. The local warming coefficients are normalized at the grid level, however scaling up resulted in the loss of data for oceans and landmasses not included in our model, so the average warming across all regions for $1^\circ$C global warming is approximately  $1.12^\circ$C. Estimates for the local warming coefficient for each region are in SI Appendix, Table S1.

We calculated vulnerability $v_i$ using data from the INFORM Global Risk Index 2020 for each country \cite{inform2020GRI}, by calculating the geometric mean of scores (out of 10) under the `Vulnerability' and `Lack of coping capacity' dimensions from the dataset. (We did not use data from the `Hazard and Exposure' dimension to avoid double counting, since this dimension is indirectly accounted for in the local warming coefficient.). We then calculated average scores across all countries in a region, omitting countries for which there is no data. Regional vulnerability estimates are in SI Appendix, Table S1.

We assumed climate change impacts for a given region, $f_i(T_i)$ could be broken down into impacts arising from local warming, $h(T_i)$, and impacts arising from the vulnerability of that region, $v_i$, so
\begin{equation}
    f_i(T_i) = h(T_i)\, \frac{v_i}{10}
\end{equation}
where $v_i/10$ represents normalized vulnerability, and $h(T_i)$ the loss associated with warming of $T_i$ degrees, where, 
\begin{equation}
    h(T_i) = s(T) m_i\,T.
    \label{eq:sT}
\end{equation} 
Here $s(T)$ is used to scale impacts with temperature and is assumed to vary with temperature. We used four data points from Figure 4a in \cite{waidelich2024climate} on global GDP impacts based on `annual temperature only' as point estimates for $s(T)$, for $T = 0.84, 1.5, 2$ and $3^\circ$C of global warming by solving
\begin{align*}
     G_W(T) &=  \sum_i \frac{p_i\, h(T_i) (v_i/10)}{P_W}\\
            &= \sum_i \frac{p_i\, s(T)\, m_i\,T\, (v_i/10)}{P_W}.
\end{align*}
$G_W(T)$ is the global GDP loss for $T$ degrees Celsius global warming, $p_i$ is the population size for region $i$ and $P_W$ the total world population. We then fitted parameters $a$, $b$ and $c$ in $s(T) = a - b\exp{(-cT)}$ to the four calculated values to get a smooth curve for $s(T)$ (fitted coefficients $a = 0.07$, $b = 0.23$, $c = 1.47$), noting that that the marginal impact of temperature increase was greater at lower temperature anomalies \cite{waidelich2024climate}. Fitted function $s(T)$, converted to percentage by multiplying $s(T)$ by 100, is shown in SI Appendix Fig.~ S5$A$, with impacts for each region, $f_i(T_i)$ in SI Appendix Fig.~ S5$B$.

Following \cite{punnavajhala2025implications}, we use survey data from \cite{andre2024globally} to estimate the initial fraction of mitigators in each region. We calculate country averages for the number of people who respond `yes' (corresponding to a value of 1) to the question asking whether they are `willing to contribute 1\% of their income to fight global warming', or, if not, respond `yes' (corresponding to a value of 0.5) to whether they are `willing to contribute less than 1\% of their income to fight global warming'. We then calculate population-weighted average (lying between 0 and 1) for each region, estimates are in SI Appendix, Table S1.

We follow the steps in \cite{punnavajhala2025implications} and use survey data from \cite{andre2024globally} to estimate the strength of social norms. We average response scores for the questions asking people whether they believe other people in their country should `try to fight global warming' (1 for `yes' and 0 for `no'), and what fraction of other people they believe are willing to contribute 1\% of their income to fight global warming. These are used to calculated population-weighted average norm strengths for each region, where the scores lie between 0 and 1 (estimates in SI Appendix, Table S1).

Historical emissions data are from \cite{guetschow2025}. For future emissions we follow \cite{punnavajhala2025implications} and assume growth according to an increasing, saturating worst case trajectory, where each regions emissions are in proportion to their future energy demand, so for region $i$,
\begin{equation}
    \epsilon_i(t) = \epsilon_i(2020) + \frac{(t-2020)\epsilon_{{\max}_i}}{t-2020+s}
    \label{eq:emissions}
\end{equation}
where $\epsilon_i(2020)$ denotes emissions for region $i$ in 2020, $t$ is the year, $\epsilon_{{\max}_i}(t)$ is the saturating value, and $s$ the half-saturating constant. Future worst case emissions are in SI Appendix Fig.~S3.

\subsection*{Calibration and simulation}
Parameters $\kappa_0$, $c_0$ and $\delta_0$ along with scaling for the initial fraction of mitigators in each region,  $\hat{x}_0$, were calibrated using Approximate Bayesian Computation (ABC) \cite{thommes2014examining} by filtering through parameter estimates giving the lowest 5\% error from a sample of 50,000 parameter sets. Parameters were calibrated to the RCP 4.5 emissions scenario \cite{pachauri2014climate}. Input ranges and median parameter estimates for three scenarios are in SI Appendix, Table S4. The model differential equations were simulated in Matlab R2021a using ode45. The model is run from 1800 to 2200, with social dynamics beginning in 2020.

\section*{Data availability}
{Data for the Shared Socioeconomic Pathways is from \url{https://tntcat.iiasa.ac.at/SspDb} and data for the Representative Concentration Pathways from \url{https://tntcat.iiasa.ac.at/RcpDb}. Data for historical carbon emissions data is from \url{https://zenodo.org/records/15016289}, and energy cost forecast data from \url{https://doi.org/10.1016/j.joule.2022.08.009}. Data on climate impacts on income are from Figure 4a of \cite{waidelich2024climate} at \url{https://doi.org/10.1038/s41558-024-01990-8}, vulnerability to climate impacts is from \url{https://data360.worldbank.org/en/indicator/INFORM_OVRL?recentYear=false&year=2020}, and location specific warming from \cite{farley2026climate}. Survey data used for initial fraction of mitigators and strength of social norms is downloaded from \url{https://dataverse.iza.org/file.xhtml?fileId=234&version=5.0}.
\section*{Code availability}
Replication code is available at  \url{https://zenodo.org/records/21288303?preview=1&token=eyJhbGciOiJIUzUxMiJ9.eyJpZCI6ImExMTQ2OGY3LWVlZmItNGJhMC1iYmI3LWMwMzIzMGJmOWI3MyIsImRhdGEiOnt9LCJyYW5kb20iOiJiMTU3YTJlMjZiYTc3NjIzMDNlY2U3YjBhY2ZmNmM5YSJ9.LUb0XeWTCwSoW0dVuXrhYJLsPYXoMdZCtSrIgsJ3vuA6DJKtvywg_yFunQLgx9zweekeSRV88TLy2HbqUPd_UQ}

\printbibliography
\clearpage

\appendix
\renewcommand{\thefigure}{S\arabic{figure}} 
\setcounter{figure}{0}

\renewcommand{\thetable}{S\arabic{table}} 
\setcounter{table}{0}

\section*{Supplementary Information}

\subsection*{Supplementary Tables}

\begin{table}[h]
    \centering
       \caption{Empirical estimates of social parameters, initial fraction of mitigators ($x_0$), strength of social norms ($\delta_0$), impact parameters, vulnerability ($v$) and local warming ($m$) in  Asia (ASIA), Latin America (LAM), the Middle-east and Africa (MAF), Organization for Economic Co-operation and Development (OECD) countries and the Reforming Economies of Eastern Europe and the former Soviet Union (REF).}
   \resizebox{\textwidth}{!}{ 
    \begin{tabular}{|l|p{3cm}|p{3cm}|p{3cm}|p{3cm}|}
         \hline
         Region& Initial fraction of mitigators ($x_0$)&Strength of social norms ($\delta$) & Vulnerability ($v$) & Local warming ($m$) \\
         \hline
         ASIA &0.76&0.66 &4.17 & 1.09\\
         LAM &0.73 &0.67&3.72&0.97\\
         MAF & 0.72&0.60&5.50&0.90\\
         OECD & 0.60&0.58&2.00&1.14\\
         REF & 0.54&0.56&3.23& 1.51\\
         \hline
    \end{tabular}}
    \label{tab:main_params}
\end{table}

\begin{table}[h]
    \centering
      \caption{Functional forms used in climate model equations. Parameter descriptions and values are in SI Appendix, Table \ref{appendix_climate_parameters}.}
    \resizebox{\textwidth}{!}{ 
    \begin{tabular}{|l|l|p{3cm}|}
    \hline
        Process & Functional form & Notes \\
        \hline
        Anthropogenic carbon emissions &  $\epsilon(t) = \epsilon_{2020} + \frac{(t-2020)\epsilon_{\max}}{(t-2020) + s}$ & (see `Emissions' under `Parameterization')\\
        \hline
        Photosynthesis & $P(C_{at},T) = \begin{cases} k_pC_{ve0}k_{MM}\Bigg( \frac{p\text{CO}_{2a}-k_c}{K_M+p\text{CO}{2a}-k_c}\Bigg)\Bigg( \frac{(15+T)^2(25-T)}{5625}\Bigg) &\text{when } p\text{CO}_{2a}\ge k_c \\ &\text{ and} -15\le T \le2 5
    \\
    0 &\text{otherwise}
    \end{cases}$ & Michaelis-Menten kinetics,\newline optimal photosynthesis at $T=2$ \\
    Ratio of molecules of CO$_2$ in atmosphere & $p\text{CO}_{2a} = \frac{f_{gtm}(C_{at}+C_{at0})}{k_a}$ & -\\
    \hline
    Plant respiration & $R_{veg}(T,C_{veg}) = k_rC_{veg}k_Ae^{-\frac{E_a}{R(T+T_0)}}$ & - \\
    Soil respiration & $R_{so}(T,C_{so}) = k_{sr}C_{so}k_Be^{-\frac{308.56}{T+T_0-227.13}}$ & - \\
    Plant death & $L(C_{veg}) = k_tC_{veg}$ & -  \\
    \hline
    Ocean flux & $F_{oc}(C_{at},C_{oc}) = F_0\chi \Bigg(C_at - \zeta \frac{C_{at0}}{C_{oc0}}C_{oc}\Bigg)$ & - \\
    \hline 
    Radiation flux &  $F_d = \frac{(1-A)S}{4}\Bigg(1+\frac{3}{4}\tau\Bigg)$ &  Assumes grey atmosphere approximation \\
    Greenhouse gas opacities & $\tau(\text{CO}_2) = 1.73(p\text{CO}_2)^{0.263}$ & \\
                            & $\tau(\text{H}_2\text{O}) = 0.0126(HP_0e^{-(L/RT)})^{0.503}$ &\\
                            & $\tau(\text{CH}_4) = 0.0231$ & \\
    \hline
    \end{tabular}    }
    \label{tab:climate_processes}
\end{table}

\begin{table}[h]
    \centering
    \caption{\textbf{Climate model parameters} Parameter values used in the climate component of the model, replicated from \cite{punnavajhala2025implications}.}
    \begin{tabular}{|l|l|l|l|}
        \hline
        Parameter & Description &Value& Unit  \\
        \hline
        C$_{\text{at0}}$ & initial CO$_2$ in atmosphere & 596 & GtC \\
        C$_{\text{ao0}}$ & initial CO$_2$ in ocean reservoir & 1.5 $\times$ $10^5$ & GtC  \\
        C$_{\text{veg0}}$ & initial CO$_2$ in vegetation reservoir & 550 & GtC \\
        C$_{\text{so0}}$ & initial CO$_2$ in soil reservoir & 1500 & GtC \\
        $T_0$ & initial average atmospheric temperature & 288.15 & K \\
        $k_p$ & photosynthesis rate constant & 0.184 & yr$^{-1}$ \\
        $k_{MM}$ & photosynthesis normalizing constant & 1.478 & 1 \\
        $k_c$ & photosynthesis compensation point & $29 \times 10^{-6}$ & 1 \\
        $K_M$ & half-saturation point for photosynthesis & $120 \times 10^{-6}$ & 1 \\
        $k_a$ & mole volume of atmosphere & $1.773 \times 10^{20}$ & moles \\
        $k_r$ & plant respiration constant & 0.092 & yr$^{-1}$ \\
        $k_A$ & plant respiration normalizing constant & 8.7039 $\times 10^9$ & 1 \\
        $E_a$ & plant respiration activation energy & 54.83 & J mol$^{-1}$\\
        k$_{sr}$ & soil respiration rate constant & 0.034 & yr$^{-1}$ \\
        k$_B$ & soil respiration normalizing constant & 157.072 & 1 \\
        k$_t$ & turnover rate constant & 0.092 & yr$^{-1}$ \\
        $C$ & specific heat capacity of Earth's surface & 4.69 $\times 10^{23}$ & J K$^{-1}$ \\
        $a_E$ & Earth's surface area & 5.101 $\times 10^{14}$ & m$^2$ \\
        $\sigma$ & Stefan-Boltzmann constant & 5.67$\times 10^{-8}$ & W $m^{-2}K^{-4}$ \\
        $L$ & latent heat per mole of water & 43.655 & mol$^{-1}$ \\
        $R$ & molar gas constant & 8.314 & J mol$^{-1}$ K $^{-1}$ \\
        $H$ & relative humidity & 0.5915 & 1 \\
        $A$ & surface albedo & 0.225 & yr$^{-1}$ \\
        $S$ & solar flux & 1368 & Wm$^{-2}$ \\
        $\tau(\text{CH}_4)$ & methane opacity & 0.0231 & 1 \\
        $P_0$ & water vapor saturation constant & 1.4 $\times 10^{11}$ & Pa \\
        $F_0$ & ocean flux rate constant & 2.5 $\times 10^{-2}$ & yr$^{-1}$ \\
        $\chi$ & characteristic CO$_2$ solubility & 0.3 & 1 \\
        $\zeta$ & evasion factor & 50 & 1\\
        \hline
    \end{tabular}
    \label{appendix_climate_parameters}
\end{table}

\begin{table}[ht]
\centering
\caption{\textbf{Assumed and calibrated parameter values.} Assumed ranges/values for parameters (from \cite{punnavajhala2025implications}) and median filtered values calibrated to Representative Concentration Pathways emissions under three scenarios using Approximate Bayesian Computation (see Methods).}
\resizebox{\textwidth}{!}{ 
\begin{tabular}{|l|c|c|c|c|c|}
\hline
Parameter & Assumed value/range & {Calibrated/calculated value} & Unit \\
\hline
Limiting value for emissions ($\epsilon_{\text{max}}$) &50 & - &GtC/year \\
Half-saturation constant for emissions ($s$) & 80 & - & -\\
Scaling for initial proportion of mitigators ($\hat{x}_0$) & [0.01, 1]    &0.52& -   \\
Scaled social learning rate ($\kappa_0$)         & [0.01, 0.8]   &0.13& -   \\
Relative climate impact cost ($c_0$)   & [0.5, 15]&7.81&    -             \\
Relative norm strength ($\delta_0$)  & [0.01, 2.5]&0.63&   -  \\
\hline
\end{tabular}
}

\label{tab:params}
\end{table}

 \FloatBarrier
% \clearpage

\subsection*{Supplementary Figures}

\begin{figure}[htbp]
\centering
\includegraphics[width=0.75\textwidth]{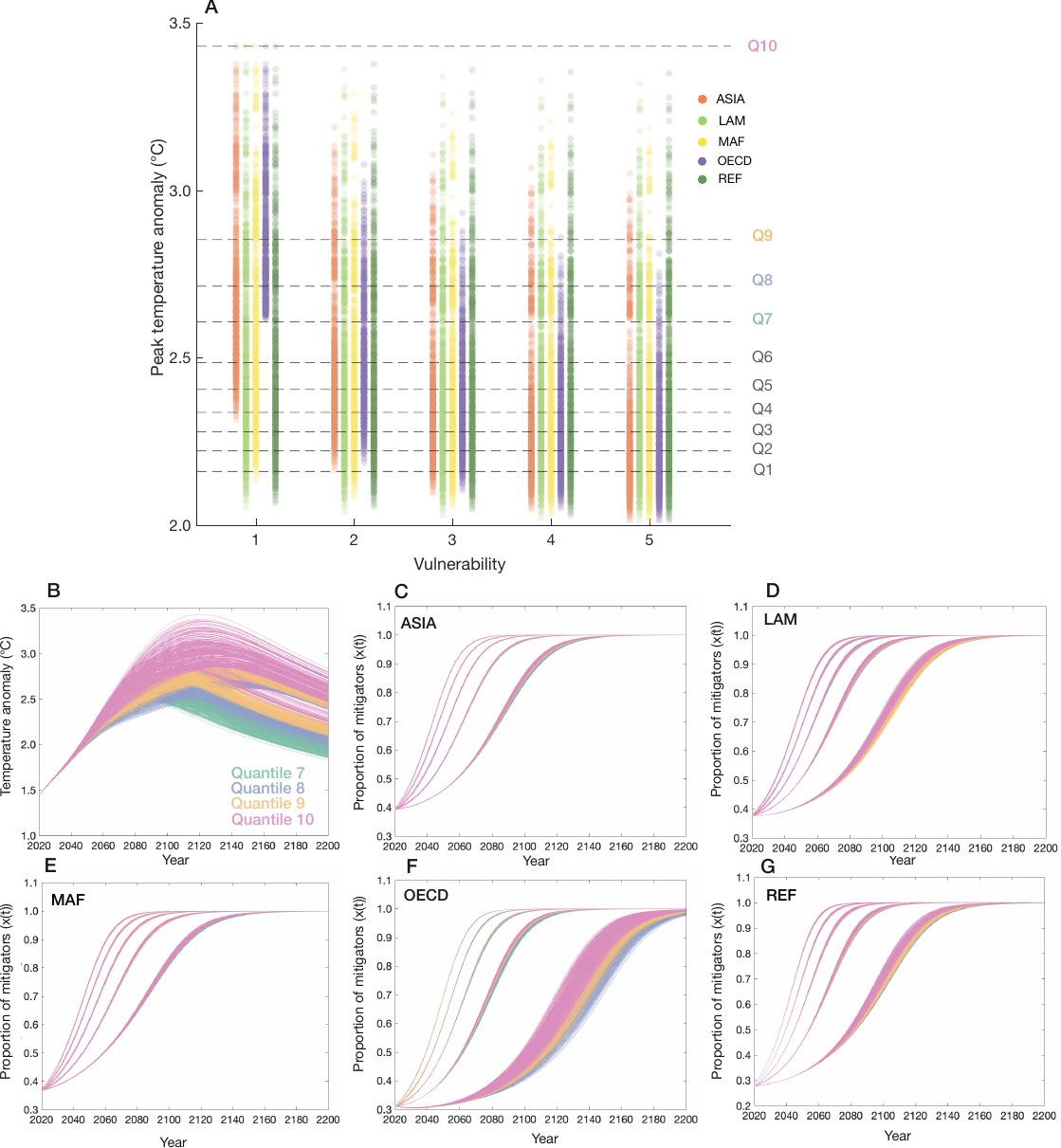}
\caption{Vulnerability in OECD drives shift in temperature pattern. ($A$) shows the distributions of peak temperatures for integer vulnerability values for each region under $2^5 = 3125$ possible permutations. The five regions shown are Asia (ASIA), Latin America (LAM), the Middle-east and Africa (MAF), Organization for Economic Co-operation and Development (OECD) countries and the Reforming Economies of Eastern Europe and the former Soviet Union (REF). Horizontal lines mark the quantiles of peak temperatures. The time evolution of temperature ($B$) and mitigation support in each region ($C$)-($G$) for vulnerability permutations that result in temperatures in quantiles 7 (green), 8 (blue), 9 (orange) and 10 (pink) from ($A$).}
\end{figure}

\begin{figure}
    \centering
    \includegraphics[width=0.9\textwidth]{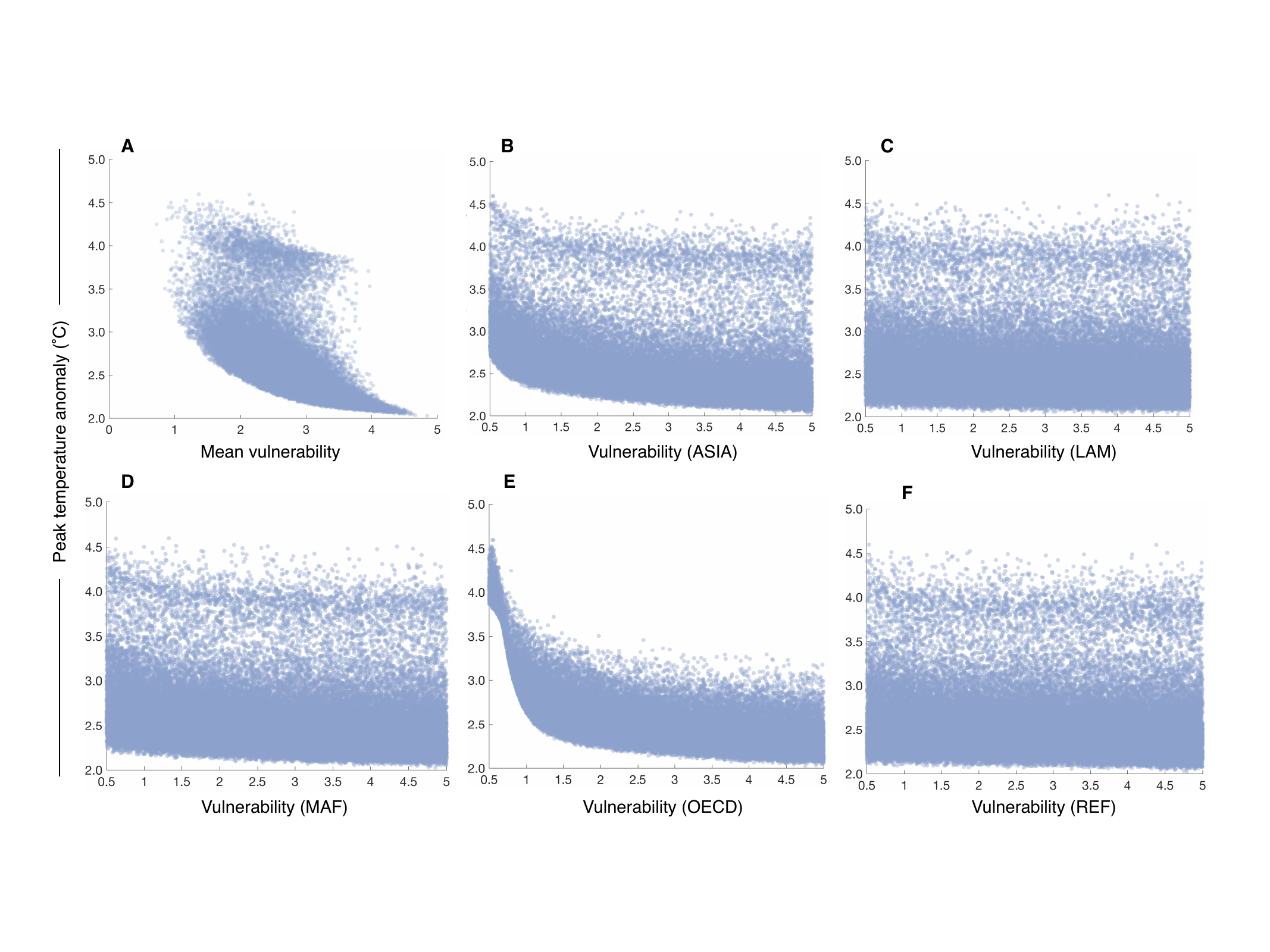}
    \caption{Bifurcation in peak temperature driven by OECD vulnerability. Peak temperature recorded for different values of average vulnerability across regions ($A$), and for vulnerability in each region, Asia (ASIA), Latin America (LAM), the Middle-east and Africa (MAF), Organization for Economic Co-operation and Development (OECD) countries and the Reforming Economies of Eastern Europe and the former Soviet Union (REF), ($B$)--($F$), respectively, when sampled randomly from a uniform distribution on [0.5,5], with $N = 50,000$.}
    \label{fig:placeholder}
\end{figure}

\begin{figure}
    \centering
    \includegraphics[width= 0.8\textwidth]{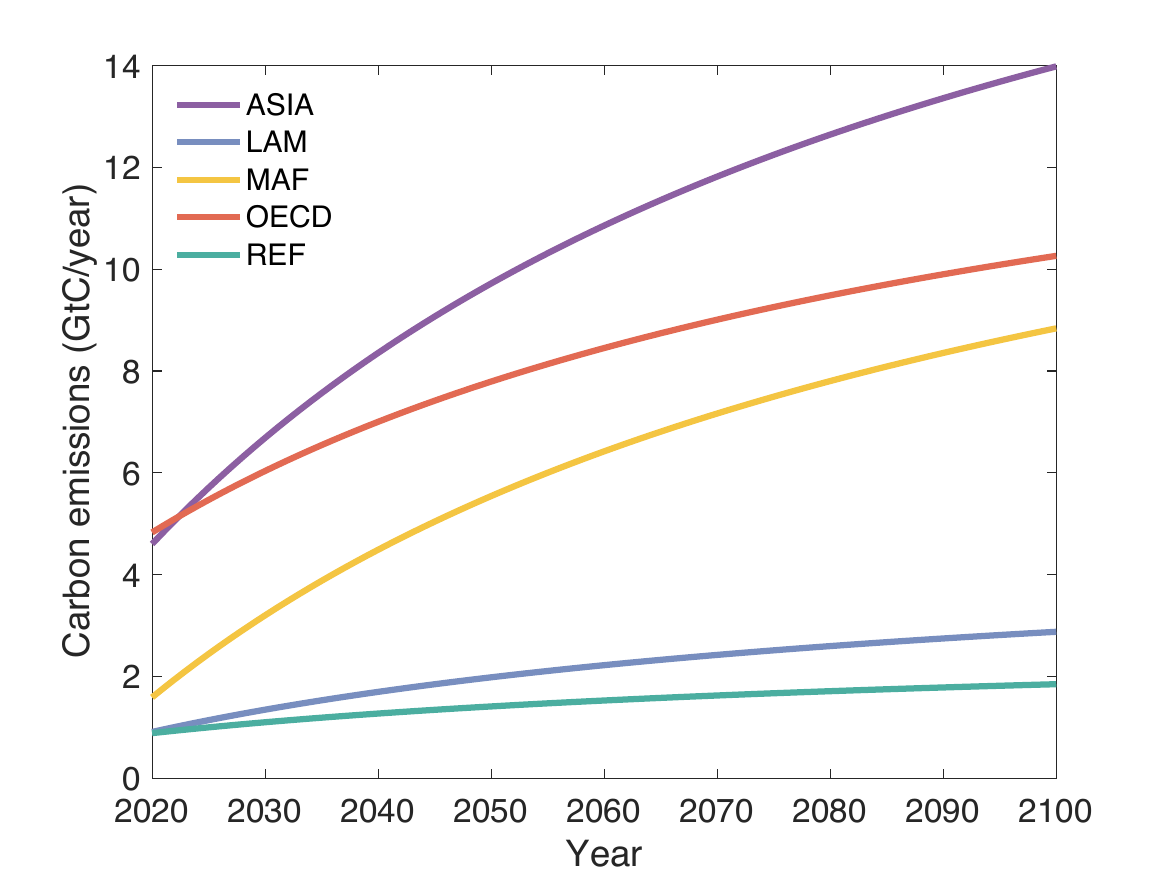}
    \caption{Worst case projected future emissions for the five regions, Asia (ASIA), Latin America (LAM), the Middle-east and Africa (MAF), Organization for Economic Co-operation and Development (OECD) countries and the Reforming Economies of Eastern Europe and the former Soviet Union (REF), from Equation (9)).}
    \label{fig:SI_emissions}
\end{figure}

\begin{figure}
    \centering
    \includegraphics[width= 0.8\textwidth]{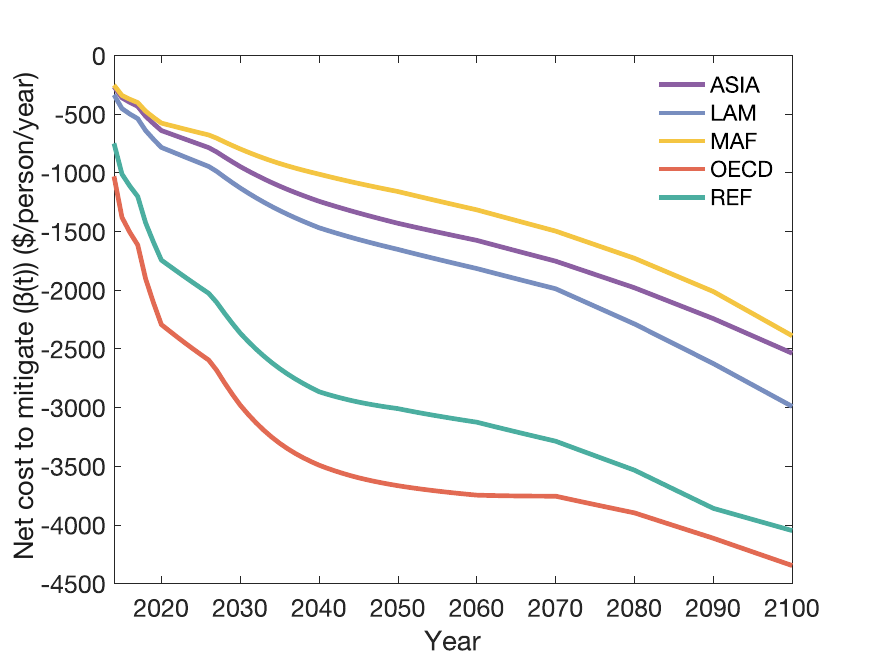}
    \caption{Projected renewable energy costs under the `fast transition' in  \cite{way2022empirically} multiplied by future energy demand for each region under Shared Socioeconomic Pathway  SSP2 from \cite{bauer2017shared}. The five regions shown are: Asia (ASIA), Latin America (LAM), the Middle-east and Africa (MAF), Organization for Economic Co-operation and Development (OECD) countries and the Reforming Economies of Eastern Europe and the former Soviet Union (REF). Horizontal dashed lines represent LCOEs for coal-powered electricity (based on \cite{way2022empirically}), multiplied by the global average energy demand.} 
    \label{fig:SI_beta}
\end{figure}

\begin{figure}
    \centering
    \includegraphics[width=\textwidth]{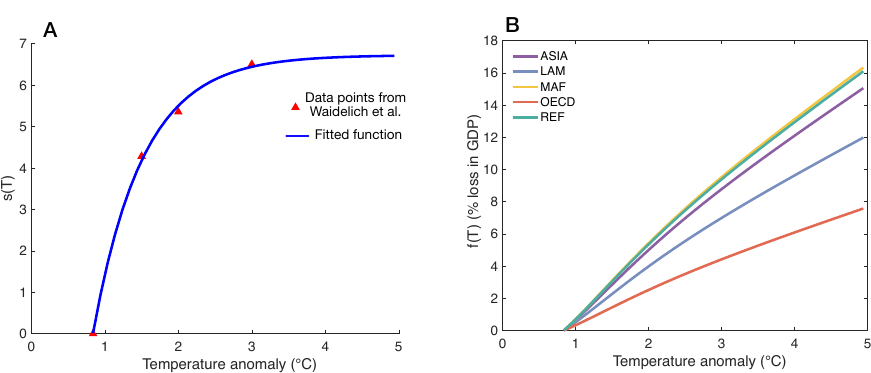}
    \caption{($A$) Fitted function $s(T)  =  a - b\exp{(-cT)}$, converted to percentage, used to scale climate impacts with temperature in Equation (\ref{eq:sT}), ($B$) cost of climate change impacts, $f(T)$ in \% loss in GDP for each of the five regions with global temperature anomaly (in $^\circ$C on the x-axis). The five regions are Asia (ASIA), Latin America (LAM), the Middle-east and Africa (MAF), Organization for Economic Co-operation and Development (OECD) countries and the Reforming Economies of Eastern Europe and the former Soviet Union (REF).}
    \label{fig:SI_fT}
\end{figure}

\end{document}